# Radial structure of vorticity in the plasma boundary of ISTTOK tokamak

B. Gonçalves [1], I. Henriques[1], C. Hidalgo [2], C. Silva [1], H. Figueiredo[1], V. Naulin[3], A. H. Nielsen[3], J.T. Mendonça[1]

[1] *Instituto de Plasmas e Fusão Nuclear, Instituto Superior Técnico, Universidade de Lisboa, 1049-001 Lisboa, Portugal.*

[2] *Euratom-Ciemat, Madrid, Spain*

[3] *Technical University of Denmark, Department of Physics, 2800 Kgs. Lyngby, Denmark*

**Abstract**

The first experimental measurement of vorticity and vorticity flux in a fusion device is an important achievement since vorticity plays a key role in the transport of energy and particles in plasmas and fluids. The measurements were performed in the plasma edge of the small tokamak ISTTOK, with an array of Langmuir probes, specifically designed for the purpose, allowing for the first time a direct comparison with theoretical models. The experimental results presented in this paper show that the vorticity flux feeds into the shear flow in the tokamak plasma edge region. The Probability Distribution Function of the vorticity exhibit fat tails with a q-Gaussian shape typical of a non-equilibrium process. Self-similarity in the probability distribution function of several parameters, including vorticity and vorticity flux, is also observed in ISTTOK and indicates that there is no morphological change in the coherent structures in the plasma boundary region and that the fluctuations in the Reynolds stress, vorticity and vorticity flux can be described by a probability distribution that tends to a universal shape.

The radial transport of particles and heat in the scrape-off layer (SOL) of magnetically confined plasmas is generally found to be turbulent and dominated by the motion of field aligned filaments, appearing as blobs in the plane perpendicular to the magnetic field, with a radial velocity component being a significant fraction of the acoustic speed. This turbulent radial transport is believed to be the cause of the experimentally observed broad particle density profiles [1,2,3,4], and large relative fluctuation levels throughout the SOL. These features strongly influence the level of plasma-wall interactions. The radial motion of a pressure perturbation initially at rest develops a dipolar structure in vorticity and electrostatic potential fields forming a radial propagating blob, in blob models of the interchange instability type [5]. The blob accelerates and develops a steep front and a trailing wake (the latter is a characteristic feature of all experimental probe measurements). The importance of energy and moment transfer between flows and turbulence in fusion plasmas is a topic that has increased in importance recently [6,7,8,9] and the simultaneous existence of multiple scales in turbulence and fluctuations was observed [10,11]. Vortex formation, in regard to structure formation, transport phenomena and turbulent flows in plasmas have been attracting much attention, and there is an increasing necessity to measure the vorticity experimentally, which is a local measure of the circulation of the velocity field at every point in the plasma fluid. Vorticity is a primary physical quantity in fluid plasma equations [12,13,14]. Vorticity is known to play a key role in the transport of energy and particles in plasmas and fluids [15]. The common feature of these models, the rate of change of vorticity, has its origin in the divergence of the plasma polarization current that provides the main perpendicular part of the charge balance in the quasineutral plasma. Therefore, measuring vorticity is essential for the validation and quantitative understanding of these models. To measure it reliably has led to significant effort in fluids studies, where hot-wires and laser anemometry have been used since long, allowing to measure three-dimensional vorticity [16,17], coherent structures [18], turbulent energy and temperature dissipation rates [19]. In fluids, vorticity is constructed from velocity field measurements using, for example, particle image velocimetry [20]. This construction is prone to inherent errors in the numerical schemes used in calculating an estimate of the curl of the velocity field.

Assuming that the plasma stream function is the electrostatic potential, which can be measured directly by suitable Langmuir probes, gives a unique advantage to the measurement of vorticity in plasmas over the measurement of vorticity in complex fluid dynamics. In spite of the strong scientific interest, experiments dealing with the dynamics of vorticity are rare. In [21] experiments on the propagation and structure of plasma filaments are described as they move across a magnetic field in a gas of neutral particles. Probe arrays were used to observe the characteristic mushroom shape and the internal electrostatic structure of the blob. Recent vortex observations were made in Argon plasmas in the High-Density Plasma Experiment (HYPER-I) [22,23,24] using directional Langmuir probes [25] to measure the flow vector field. Recently, a 7-tip Langmuir probe for vorticity measurements has been used at the Large Plasma Device at UCLA [26,27] to the characterization of vortices in the Kelvin–Helmholtz instability and to characterize coherent structures driven by near steady-state shear-flows. Also experiments in the Controlled Shear Decorrelation Experiment, a 2.8 m long linear helicon plasma device, [28] determined that collisional electron drift wave turbulence generates drift wave packet structures with density and vorticity fluctuations (measured with a 3x3 probe array [29]) in the central plasma pressure gradient region of a linear plasma device. Nonlinear energy transfer measurements and time-delay analysis confirm that structure absorption amplifies the sheared flow. In this work it is stated that similar mechanisms likely operate at the edge of confined toroidal plasmas and should lead to the amplification of sheared flows at the boundary of these devices as well.

In a turbulence the formation and motion of vortical structures gives rise to a density flux, due to the density perturbation associated with the structures. It was shown that the flux connected with higher order drifts (higher than ExB drift), as the nonlinear polarization drift, is directly linked to the motion of the vortex structures [30]. The polarization drift even if it is small compared to the ExB drift, leads to a transport component which can be large. Most important, it causes most of the large transport events. This effect is further related to the Reynolds stress, which is made responsible for the appearance of radial electric fields in the plasma and generation of poloidal flows. The first evidence of the importance of Reynolds stress in turbulent fusion plasmas was found in ISTTOK tokamak using a Langmuir probe array [31]. Moreover, the vortex structures not only transport particles, they carry polarisation charge and can be able to organize large-scale potential differences inducing flows perpendicular to the background density gradient. These zonal flows [32] play an important role in the formation of transport barriers.

In this paper the measurements obtained from a specifically designed array of Langmuir probes, are presented. Experiments were carried out in ISTTOK [33], is a small tokamak (major radius $R = 0.46$ m, minor radius $r = 0.085$ m, toroidal magnetic field $B \sim 0.5$ T, $T_{e0} \sim 100\text{-}150$ eV, $n_{e0} \sim 6.5 \times 10^{18}$ m$^{-3}$ and plasma current $I_p = 4\text{-}6$ kA) of circular cross section supporting AC operation (plasma current inverted periodically, in a time scale of typically 30 ms [33]) with a total duration up to 1s. The Deuterium plasma main parameters for the set of discharges under consideration in this paper were $I_P \sim 4$kA flat top and $n_e \sim 3.5\text{-}4 \times 10^{18}$ m$^{-3}$. The use of AC discharges allows for a bigger set of experimental results for each radial position an in this work three comparable cycles were obtained per discharge. Edge parameters were digitized at 2 Mega Samples Per Second (MSPS) from a 3-5 ms time window during each semi-cycle discharge flat top (~20 ms). Plasma profiles and turbulence have been investigated using the probe head, located on the equatorial plane of the device. It consists of two parallel arrays of Langmuir probes separated by Δr~3 mm (Fig. 1) allowing the simultaneous investigation of the radial structure of fluctuations on vorticity, Reynolds stress and turbulence in the plasma boundary region. Measurements were taken at different radial positions, both in the edge (r < $a_{limiter}$) and in the scrape-off layer (SOL) (r > $a_{limiter}$) on a shot by shot basis. For each position the first three positive cycles of the AC discharge were used for the analysis. Two tips of each set of three probes, aligned perpendicular to the magnetic field and separated poloidally (Δθ~5 mm), were used to measure fluctuations of the poloidal electric field $\tilde{E}_\theta$, as deduced from the floating potential ($V_f$) and neglecting electron temperature fluctuation effects [i.e., $\tilde{E}_{\theta 1} = (\tilde{V}_{f4} - \tilde{V}_{f5})/\Delta\theta$ and $\tilde{E}_{\theta 2} = (\tilde{V}_{f5} - \tilde{V}_{f6})/\Delta\theta$]. When temperature fluctuations are of significance, the ion saturation current fluctuations contains a contribution from temperature fluctuations and the relation between the floating potential and the plasma potential also depends on the temperature. Recent studies using GEMR gyro-fluid simulations [34] raised some concerns regarding the reliability of Langmuir probe measurements for plasma-turbulence investigations. These simulations have shown that Ion-saturation current measurements turn out to reproduce density fluctuations quite well. However, fluctuations in the floating potential, are strongly influenced by temperature fluctuations and, hence, are strongly distorted compared to the actual plasma potential. These results suggest that interpreting floating as plasma-potential fluctuations while disregarding temperature effects is not justified near the separatrix of hot fusion plasmas. Here, floating potential measurements led to corrupted results on the E × B dynamics of turbulent structures in the context of, e.g., turbulent particle and momentum transport or turbulence characterization on the basis of density–potential phase relations. Results obtained on ISTTOK indicate that floating potential measurements by Langmuir probes overestimate the amplitude of the plasma potential fluctuations due to the influence of the electron temperature fluctuations [35], but potential fluctuations measured by Langmuir Probes and Ball Pen Probes were found to be well correlated and roughly in phase. This result has important implications, justifying the similarities in the statistical properties of the plasma parameters measured with both types of probes as well as in the phase velocity of the fluctuations and in the poloidal correlation. The frequency and wave-number spectra also do not show major differences when measured with both types of probes.

The sixth tip was biased at a fixed voltage in the ion saturation current regime $I_S$. The radial electric field was estimated from floating potential signals measured by radially separated probes (i.e., $\tilde{E}_r = (\tilde{V}_{f3} - \tilde{V}_{f2})/\Delta r$) oriented with respect to the magnetic field direction to avoid shadowing [36]. We assume the fluctuations of the ion saturation current in the SOL to be proportional to density fluctuations $\tilde{n}$ [37]. The associated turbulent flux is computed using $\Gamma_{E \times B} \cong \langle \tilde{n}\tilde{E}_\theta \rangle /B$ [38] under the assumption that the electron temperature fluctuations are negligible [39]. Particular care was taken to only use data from steady state plasmas, in order to guarantee meaningful statistics.

The principle behind the vorticity probe is the use of Langmuir probes in the stencil of discrete approximation of the Laplacian used in numerical computations. The 5 probe tips are aligned perpendicular to the magnetic field in approximately Diamond pattern, separated radially and poloidaly (fig. 1). The vorticity fluctuations were computed from the fluctuations in the ExB velocities measured by the probe following the equation

$$\widetilde{\omega} = \frac{1}{B}\left(\frac{\tilde{V}_{f2} - 2\times \tilde{V}_{f5} + \tilde{V}_{f3}}{\Delta r^2} + \frac{\tilde{V}_{f6} - 2\times \tilde{V}_{f5} + \tilde{V}_{f4}}{\Delta \theta^2}\right)$$

The flux of vorticity can also be estimated as $\langle \tilde{v}_r \widetilde{\omega}\rangle \propto \langle \tilde{E}_\theta \widetilde{\omega}\rangle/B$. The radial profiles of the Ion saturation current and floating potential are shown in Fig. 2. The floating potential becomes more negative when the probe is inserted into the plasma edge and the radial electric field changes its sign in the proximity of the limiter radius location. There is a significant gradient in the RMS fluctuation level (depicted as the error bars in the graphics) which increases radially inwards.

Figure 3a shows the radial profile of the poloidal phase velocity. The phase velocity was computed from the statistical wavenumber-frequency spectrum S(k,ω) function [40], computed from the two-point correlation technique using two floating potential signals as $v_\theta = \sum_{\omega,k}(\omega/k)S(k,\omega)/\sum_{\omega,k}S(k,\omega)$. The poloidal phase velocity presents a clear change in the propagation direction of fluctuations from ion diamagnetic direction in the outer edge of the plasma behind the limiter to electron direction inside the limiter radius. The dispersion observed in the figure is due to the radial movement of the plasma column (< 10 mm) between the different positive cycles of the AC discharges analysed. This is also the reason why, although the exact location of the limiter it is known, it is marked in the graphics as a shaded area in the position -10 mm to 0 mm. A clear shear layer is also observed in the poloidal velocity near the limiter position. The existence of a naturally occurring shear layer near the Last Closed Flux Surface (LCFS) was observed previously in tokamaks [41,42], stellarators [43] and reversed field pinches [44]. The particle flux is depicted in Fig 3b. Its profile shows a maximum in the region just inside the LCFS as typically observed in fusion devices. The profiles of the skewness and kurtosis of the particle flux show a value closer to a Gaussian in the region of the shear layer (Figures 3c and 3d). This reduction is expected as result of the decorrelation of the transport events due to the shear effect. Although there are not many results showing the radial profile of turbulent transport at the plasma edge, measurements at RFX have shown a maximum of the turbulent transport close to the shear layer and a reduction at the plasma edge [45]. Poloidal asymmetries could explain the negative transport on ISTTOK, however other interpretations are also valid as for example the formation of collective cells as the ones observed in TJ-K/HSX [46]. In [47] it was shown that electrostatic Reynolds stress term (Re) dominates the particle flux. This term was determined by Re = $\langle \tilde{v}_r \tilde{v}_\theta \rangle$ [48]. The $\langle \tilde{v}_r \tilde{v}_\theta \rangle$ term of the Reynolds stress tensor can be related to the ExB velocities, $\langle \tilde{v}_r \tilde{v}_\theta \rangle \propto \langle \tilde{E}_r \tilde{E}_\theta \rangle$, $\tilde{E}_r$ and $\tilde{E}_\theta$ being the radial and poloidal components of the electric field, respectively. where $\tilde{E}_\theta$ is the mean value of $\tilde{E}_{\theta 1}$ and $\tilde{E}_{\theta 2}$. In this way the poloidal and radial components of the electric field are estimated, approximately, at the same plasma position. The brackets (< >) denote time average over times significantly longer than correlation times. It should be noted that only in the case of poloidally homogeneous turbulence the Reynolds stress, computed as the time averaged product of fluctuating radial and poloidal velocities, and the flux surface average Reynolds stress [48,49] are equivalent. Radially varying Reynolds stress allows the turbulence to rearrange the profile of poloidal momentum, generating sheared poloidal flows. In this way the poloidal and radial components of the electric field are estimated, approximately, at the same plasma position. The computed value of $\langle \tilde{E}_r \tilde{E}_\theta \rangle$, is very similar to those deduced using the values of the fluctuating poloidal electric field measured by the inner $\tilde{E}_{\theta 1}$ or outer $\tilde{E}_{\theta 2}$ probes. Figure 4a shows the $\langle \tilde{v}_r \tilde{v}_\theta \rangle$ radial profile, which exhibits a radial gradient in the proximity of the velocity shear layer location. Similar measures were made on ISTTOK in the past showing that this mechanism can drive significant poloidal flows in the plasma boundary region [50]. Figures 4b and 4c show the radial profile of the vorticity and of the vorticity flux respectively. It was observed that the vorticity is constant in the SOL and limiter region but in the plasma edge, where the ExB shear flow is higher a larger dispersion is observed in the vorticity measurements (as well as on the measured poloidal velocities observed in fig. 3a) and the profile shows a peak at position -15 mm and inverts its sign at the position -20 mm. The observed dispersion is not unexpected as theoretical models have predicted that the magnitude of the shear layer leads to selectivity of the vorticity [51]. In this work it was shown that in the absence of the sheared flow, the axial (perpendicular to the plane) vorticity field would be rather homogeneous and isotropic while a sheared flow carries an associated constant axial vorticity that is added to the background vorticity. The vorticity flux is positive in the plasma edge and decays to close to zero (or even negative) towards the limiter region. In [52] it is shown that for ExB-dominated turbulent flows which have azimuthally invariant fluctuation statistics, the Taylor identity holds and shows that the vorticity flux is related to the turbulent Reynolds stress and Reynolds force $F_\theta^R$ exerted by the fluctuations upon the background plasma by the relation $F_\theta^R = -\nabla_r \langle \tilde{v}_r \tilde{v}_\theta \rangle = -\langle \tilde{v}_r \tilde{\omega} \rangle$ [53,54]. Both the Reynolds stress and the vorticity flux exhibit a strong gradient in the region inside the limiter. The order of magnitude of the gradient is within the expected values, meaning that the Reynolds force $F_\theta^R$ resulting from the vorticity flux amplifies the shear flow in the tokamak plasma edge region. According to the results the vorticity flux may strongly contribute to the shear flow amplification in the plasma edge region of tokamaks. These results also show that the particle flux and vorticity flux are closely related. Similar results were also observed linear plasma devices [28]. Figures 5a and 5b show the pdf of vorticity and vorticity flux, respectively, for different radial positions. The distribution becomes broader as the probes are moved into the plasma, with larger events being clearly visible in the tails of the PDF which in both cases is asymmetric and dominated by positive events. This asymmetry is clearly observed in the radial profile of the skewness (fig.5c and 5d) which is positive for all distributions. The radial profile of the skewness (Fig. 5c and 5d) and kurtosis (fig. 5e and 5f) also clearly show a strong decrease at the shear layer location where the velocity shear is higher.

Plasma density fluctuations and electrostatic turbulent fluxes measured at the scrape-off layer of the Alcator C-Mod tokamak [55] the Wendelstein 7-Advanced Stellarator [56], the TJ-II stellarator [57] and JET [57] are shown to obey a non-Gaussian but apparently universal (i.e., not dependent on device and discharge parameters) probability density distribution (PDF). It has been reported that density fluctuations are distributed according to non-Gaussian probability density functions [58,59,60,61] and that the associated turbulent fluxes seem to be approximately self-similar (i.e., invariant under rescaling) over a range of scales, larger than the characteristic turbulent scales [62]. In [63] was reported experimental evidence showing that fluctuations of the density and the flux in the SOL of tokamaks and stellarators tend to adopt a canonical shape, which acts as an attractor for the PDF. The fact that a specific nontrivial universal shape acts as an attractor for the PDF seems to suggest that emergent behavior and self-regulation are relevant concepts for these fluctuations. The results presented in [63] are consistent with radial transport being dominated by large-scale density structures (blobs). Self-similarity in the probability density function of turbulent transport in the edge plasma region, with a rescaling parameter dependent of the level of fluctuations described in [64]. Figure 6a shows the pdf of the turbulent flux normalized to its signal level (Root Mean Square, RMS) for the different radial locations. The PDFs show some degree of self-similarity. It is interesting to note that the self-similarity also holds for the PDFs of the Reynolds stress, vorticity and vorticity flux (figures 6b to 6d, respectively) when the time series are normalized to the corresponding signal level although in these cases the tales of the distribution show some scattering likely due to a reduced number of events larger than several RMS. The self-similarity observed in ISTTOK seems to indicate that there is no morphological change in the coherent structure in the plasma boundary region and that the fluctuation in these quantities can be described by a probability distribution that tends to a universal shape. The vorticity PDF (Figure 6c) is non-Gaussian, showing a fat tail typical of strongly correlated systems, and slightly asymmetrical towards the large scale positive vorticity events implying the existence of large intermittent coherent structures. Among various non-Gaussian distributions that emerge from consistent thermodynamical and statistical frameworks [65,66,67], q-Gaussians, based on the so-called non-extensive statistical mechanics introduced by Tsallis [68], are appealing for their simplicity. Many experimental measurements of distribution functions of particle systems and other physical quantities can be described by non-Gaussian distributions. Q-Gaussians have been employed in the study of probabilistic models [69], space plasmas[70], earthquakes [71] and the solar wind[72]. The q-Gaussian distribution is specified by the pdf

$$p_{qg}(x) = p_0 \left[1 - (1-q)\left(\frac{x}{x_0}\right)^2\right]^{1/(1-q)}$$

For $1-(1-q)(x/x_0)^2 \geq 0$ and $p_{qg}(x)=0$ otherwise. Figure 7 shows the probability distribution function of vorticity obtained from the experimental measurements obtained at different radial position sfor three cycle of the AC discharge. The PDFs of the vorticity exhibit fat tails with a q-Gaussian shape typical of a non-equilibrium process. Figure 7e) shows the overlap of the probability distribution function of vorticity obtained from the experimental measurements obtained at a fixed radial position for three cycle of the AC discharge and the result of a q-Gaussian fit with q=1.494. For q=1 we would be reduced to a Gaussian distribution valid in equilibrium. Figure 8 shows the radial variation of the q parameter resulting from the fit performed to the pdfs of vorticity at different radial positions. A dip is observed in the radial profile, close to the limiter position with a q closer to that of a Gaussian distribution. The existence of positive tails in pdf distribution is also interesting, because as shown in [47] a similarity exists between transport statistics, particularly for positive transport events, during ELMy H-mode and L-mode, where the transport is known to be mediated by radially propagating blob structures. Although the magnitude of the transport varies by a large factor, the results indicate a strong similarity of the underlying joint transport mechanisms.

The experimental study of vorticity is an important achievement in the plasma studies, since it is known to exist in turbulent plasmas, and allows direct comparison with theoretical models. The results show no morphological change in the coherent structures in the plasma boundary region and that the fluctuations in the Reynolds stress, vorticity and vorticity flux can be described by a probability distribution that tends to a universal shape. It was shown that the probability distribution function of the vorticity can be fitted by a q-Gaussian distribution typical of a non-equilibrium process. The similarity of transport mechanisms during ELMy H-mode and L-mode previously observed in [47] make these results also highly relevant. Future work will focus on performing polarization experiments on ISTTOK to study the effects of the ExB shear in the vorticity, the comparison of the result with numerical simulations, such as 2D numerical turbulent interchange model HESEL [73] and GBS code [74] and on performing similar measurements in larger fusion devices to allow direct comparison between L and H-mode.

**Acknowledgements**

IPFN activities received financial support from "Fundação para a Ciência e Tecnologia" through project UID/FIS/50010/2013.

**Figure Captions**

Fig. 1: a) Tips of the vorticity probe: ion saturation current ($I_{Sat}$) in red, and floating potential ($\phi_f$) in blue; and b) probe dimensions (axis: r radial, $\theta$ poloidal and $\phi$ toroidal).

Fig. 2: Average radial profiles of a) $I_S$, b) RMS ($I_S$), and radial profiles of c) $V_f$ and d) RMS ($V_f$) for the three cycle of the AC discharge observed. The profiles show the consistency of the measurements at different cycles

Fig. 3: Radial profile of a) poloidal velocity for the three cycle of the AC discharge observed; b) average ExB particle flux; c) Skewness of the ExB particle flux Probability Distribution Function and d) Kurtosis of the ExB particle Flux Probability distribution function

Fig4: Radial profiles of a) Reynolds stress; b) vorticity; and c) Vorticity flux and the corresponding error bars

Fig. 5: Probability distribution function at different radial locations of a) vorticity and b) vorticity flux and respective radial profiles of skewness and kurtosis for the vorticitiy and vorticity flux PDFs

Fig 6: Probability Distribution Function (PDF) normalized to the standard deviation of the signal fluctuations, at different radial locations, of a) ExB particle flux; b) Reynolds stress; c) vorticity; and d) vorticity flux

Fig 7: Probability distribution function of the vorticity obtained from the experimental data (scattered points) from a) the radial position -20mm (in the edge) to e) the radial position r=5 mm located in the SOL. Each graphic corresponds to a radial displacement of 5 mm. The line on figure f) indicates the q-Gaussian fit to the experimental data with a q=1.494 while the dashed line shows a Gaussian fit to the data

Fig 8: Radial profile of the q parameter resulting from the fit of a q-Gaussian to the probability distribution function of the vorticity

**Figures**

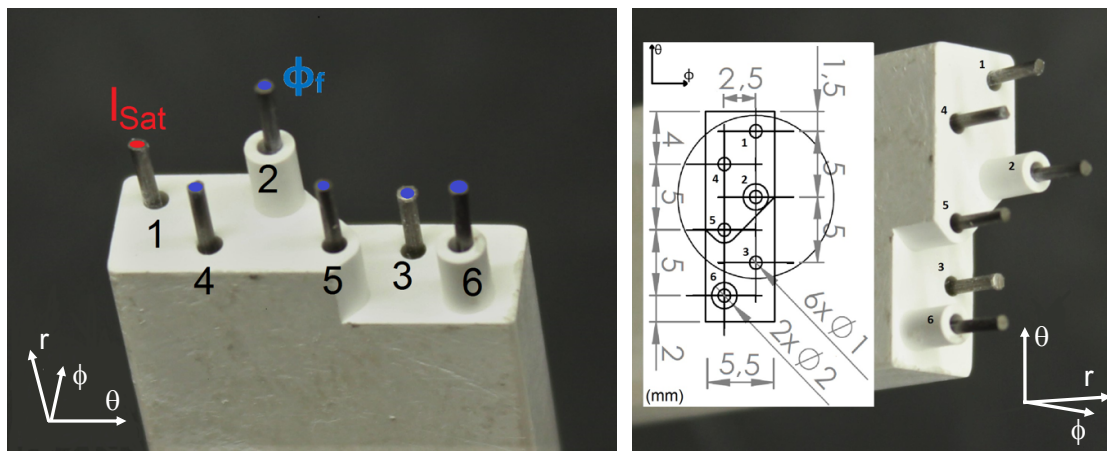

**Fig. 1**

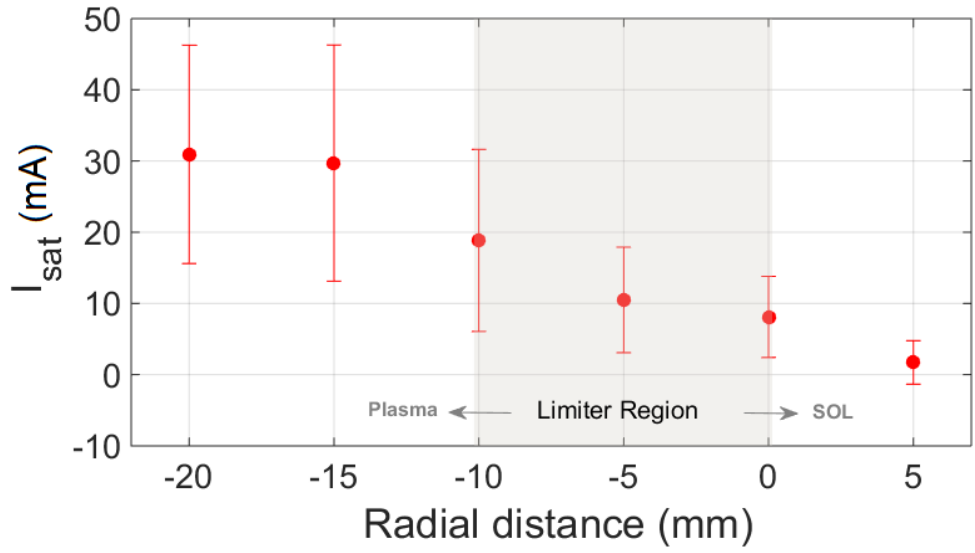

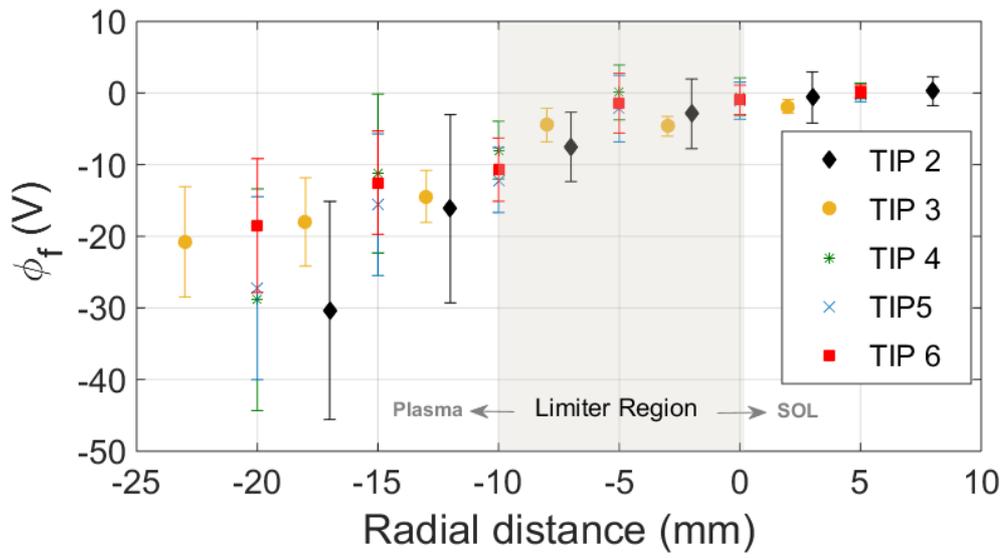

**Fig. 2**

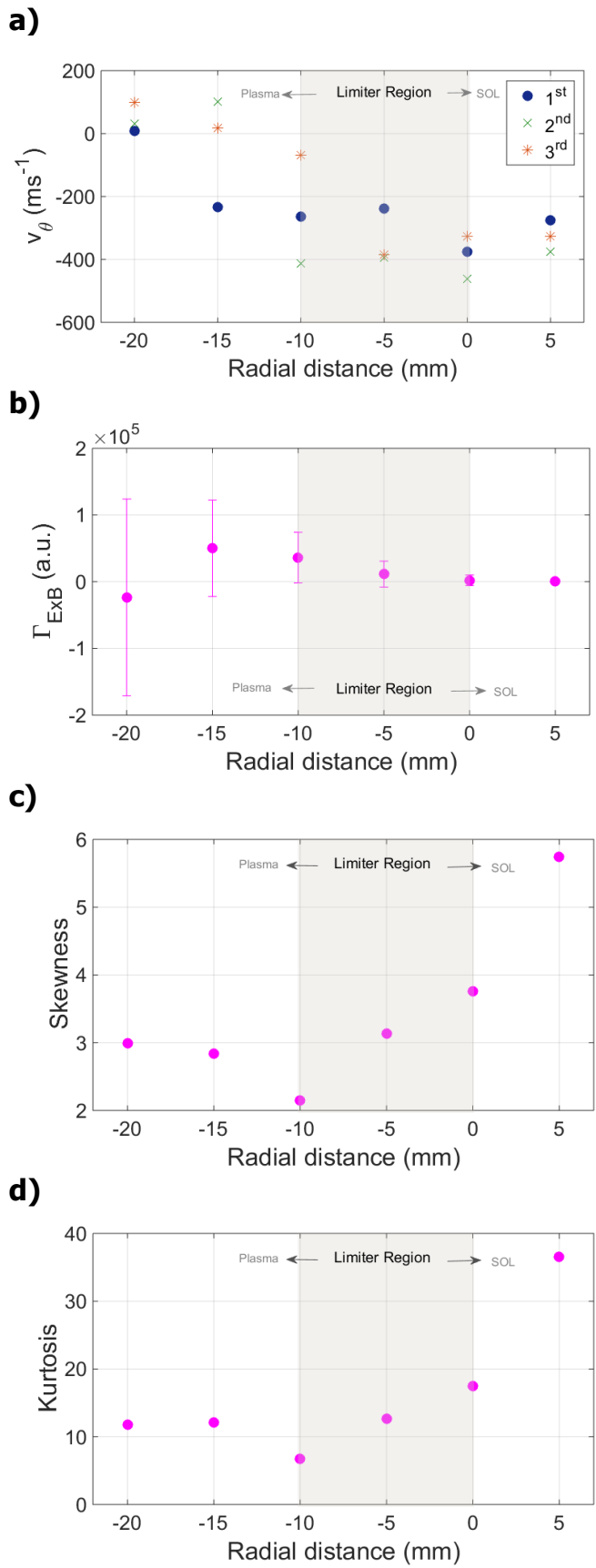

**Fig.3:**

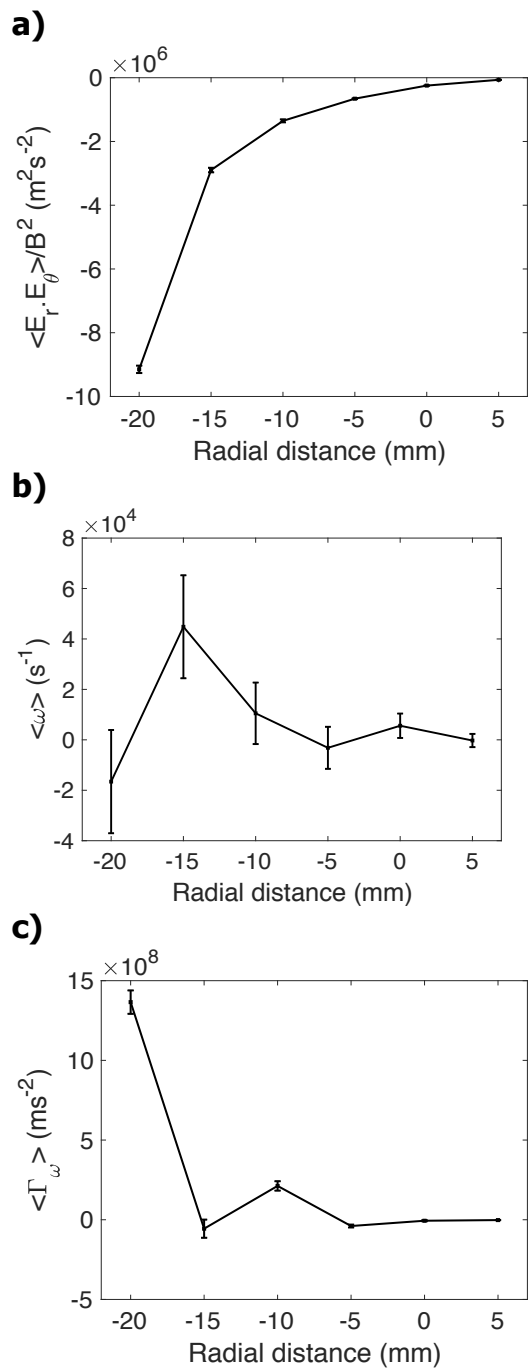

**Fig. 4:**

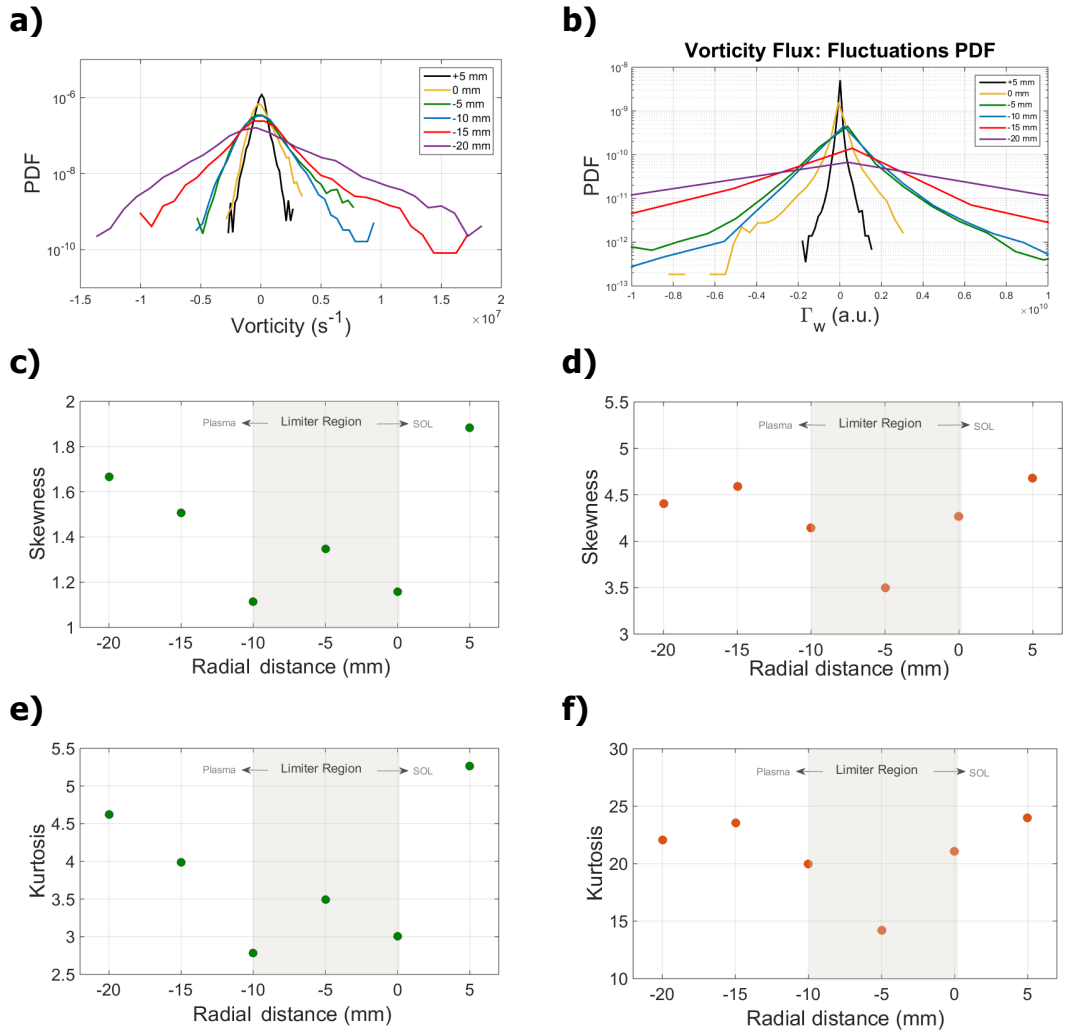

**Fig. 5**

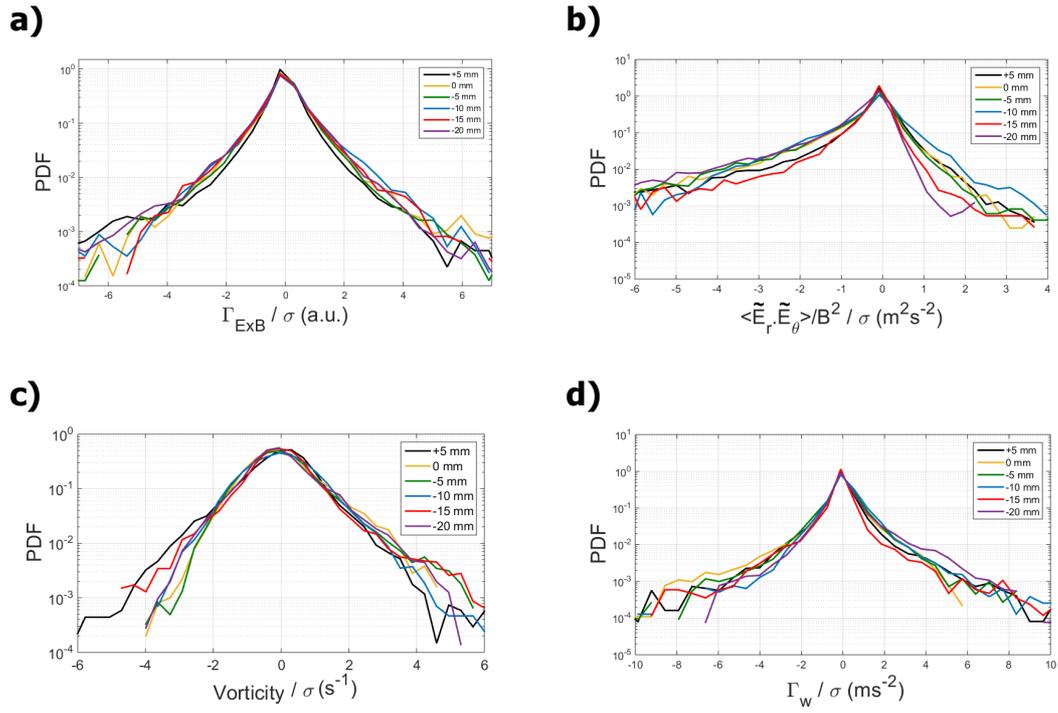

**Fig. 6**

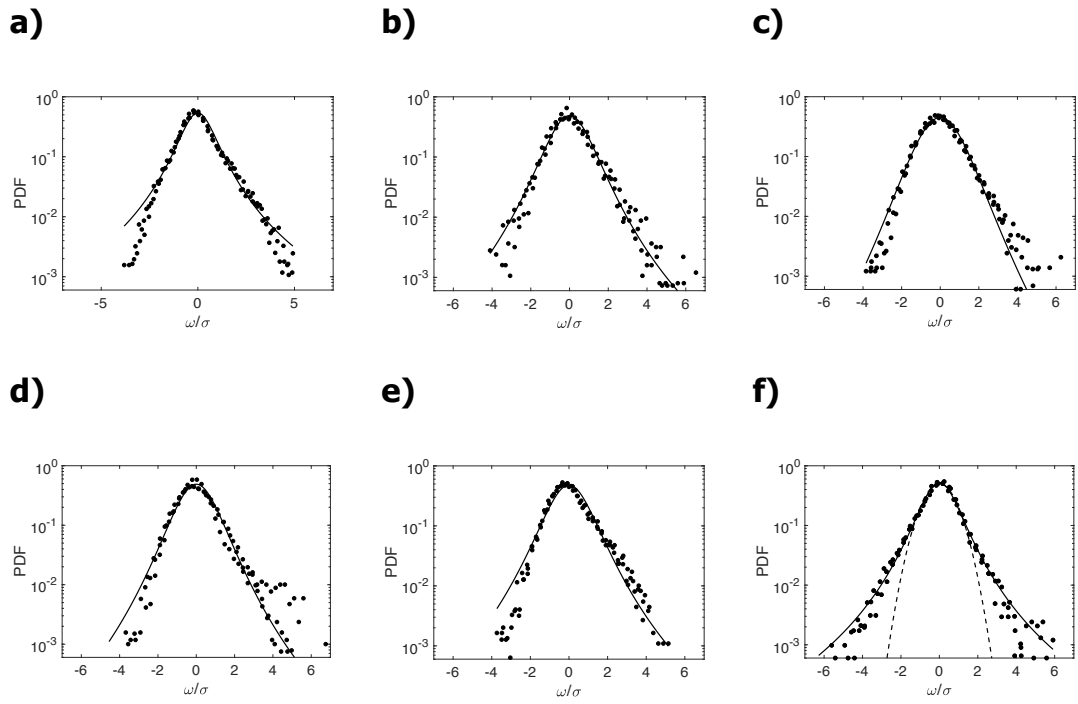

**Fig. 7**

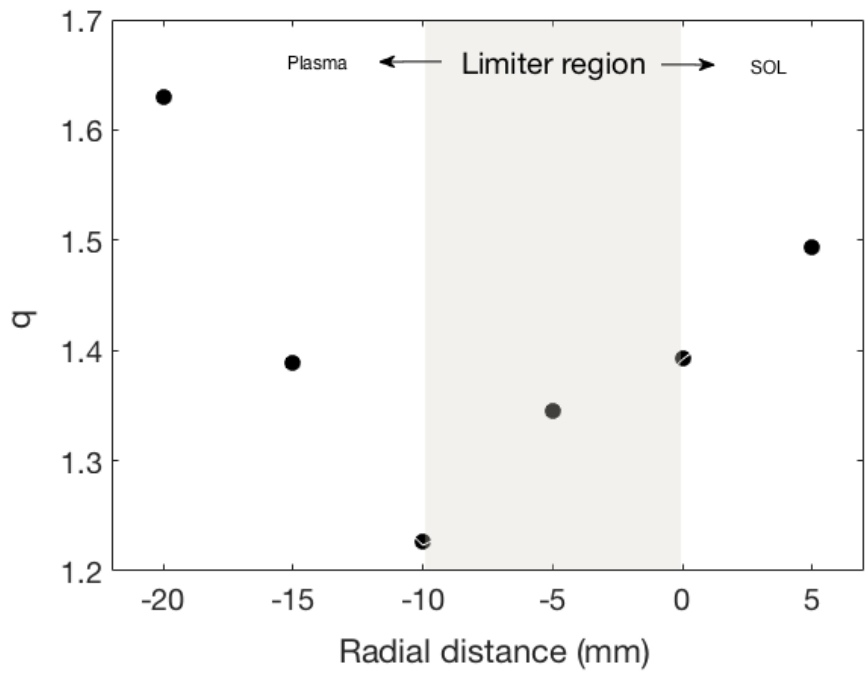

**Fig. 8**